\documentclass[aps,preprint,showpacs]{revtex4}%

\usepackage{amsfonts}
\usepackage{amsmath}
\usepackage{amssymb}
\usepackage{graphicx}%
\begin{document}
\title{DNA in nanopore - counterion condensation and coion depletion}
\author{Yitzhak Rabin}
\affiliation{Institute for Mathematics and its Applications, University of Minnesota,
Minneapolis, MN 55255, USA and Department of Physics, Bar-Ilan University,
Ramat-Gan 52900, Israel}
\author{Motohiko Tanaka}
\affiliation{National Institute for Fusion Science, Toki 509-5292, Japan}

\begin{abstract}
Molecular dynamics simulations are used to study the equilibrium distribution
of monovalent ions in a nanopore connecting two water reservoirs separated by
a membrane, both for the empty pore and that with a single stranded DNA
molecule inside. In the presence of DNA, the counterions condense on the
stretched macromolecule effectively neutralizing it, and nearly complete
depletion of coions from the pore is observed. The implications of our results
for experiments on DNA\ translocation through $\alpha$-hemolysin nanopores are discussed.

\end{abstract}
\date{December 8, 2004} 

\startpage{1}
\endpage{ }
\maketitle


Recent experiments on translocation of a single stranded (ss) DNA molecule
through an $\alpha$-hemolysin nanopore inserted in a
membrane\cite{Kasianowicz,Akeson,Meller1,Meller2,Bayley} have prompted a
number of theoretical studies\cite{Lubensky,Kholomeisky}. These studies, as
well as other models of polymer translocation through narrow
pores\cite{Yoon,Sung,Obukhov,DiMarzio,deGennes,Muthukumar,Kantor,DePablo},
focused on the polymer aspects of the problem such as the effect of
confinement-induced entropy losses on the translocation of a Gaussian chain
and on the effective friction between the chain and the narrow pore. The fact
that DNA is a charged object was considered only as far as the mechanism of
pulling by an externally applied potential gradient was concerned. While the
above theories were quite successful in reproducing the observed distribution
of translocation times\cite{Lubensky} and the molecular weight dependence of
the translocation velocity\cite{Kholomeisky}, some important questions remain
unanswered. For example, as has been noted in reference \cite{Lubensky}, the
theory overestimates (by more than an order of magnitude) the effective
pulling force on the DNA. A similar discrepancy concerning the magnitude of
the force acting on DNA has been observed in recent experiments on
nanopore-induced opening of DNA hairpins\cite{Jerome}. Furthermore, while the
models focus on the kinetics of DNA translocation through the pore,
experiments probe this process only indirectly, by monitoring the transient
blocking of the current\ of small ions through the channel. It is therefore
essential to understand the physics of these ions in confined space - their
interactions with the pore and with the DNA molecule, as well as their mutual
interactions. Notice that even though the inner diameter of the $\alpha
$-hemolysin pore is much larger than the size of a small ion, the presence of
negatively charged DNA is expected to have a dramatic effect on the
distribution of small ions in the pore.

We proceed to study the problem by molecular dynamics simulations. We take an
isolated rectangular box of three sides $56$ $\mathring{A}$, $56$
$\mathring{A}$ and $112$ $\mathring{A}$ in the x, y and z directions,
respectively. This box is separated into upper and lower compartments by a
membrane of thickness $50$ $\mathring{A}$. The center of the membrane is
permeated by a cylindrical pore of radius $7.5$ $\mathring{A}$ positioned
along the central z-axis and assumed to be neutral (we do not account for the
presence of charged and polar amino-acids at the membrane-water
interface\cite{Roux}). In the absence of DNA the simulation box is filled with
three particle species, namely, counterions, coions and water molecules that
are free to move throughout the upper and lower compartments and through the
pore, but can not permeate the volume occupied by the membrane. The
counterions and coions are spheres of radii $1.3$ $\mathring{A}$ and $1.8$
$\mathring{A}$ corresponding to ionic radii of $K^{+}$ and $Cl^{-}$
respectively, having either a positive or negative unit charge at their
centers. There are $100$ salt ions of each sign in the simulation box,
corresponding to salt concentration of about $1M$. Water molecules are
emulated by putting a sphere of radius $1.2$ $\mathring{A}$ in every $(2.9$
$\mathring{A})^{3}$ volume, typically amounting to $6800$ particles (a value
smaller than the Van der Waals radius of water was taken in order to speed up
the simulation). The electrostatic properties of water and of the membrane
have been taken into account by treating them as dielectric continua with
$\varepsilon_{w}=80$ and $\varepsilon_{m}=2$ respectively, and matching the
two dielectrics by introducing a $2.1$ $\mathring{A}$ thick transition layer
at the water-membrane boundary, across which the dielectric constant changes
linearly from the former to the latter value. Reflecting boundary conditions
are assumed at the surfaces of the rectangular box and of the membrane, and at
the pore wall.

The ssDNA is modeled as a flexible chain of $12$ repeat units each of which
contains a backbone made of a spherical bead with charge $-e$ at its center
(phosphate group) connected to a non-charged bead (sugar ring), to which an
additional non-charged bead (base) is attached. The beads are connected by
effective springs and undergo thermal fluctuations under the constraint that
the center of the macromolecule is pinned down at the pore center. The radii
of the charged, non-charged, and side-chain beads are taken as $2.1$
$\mathring{A}$, $1.8$ $\mathring{A}$ and $2.2$ $\mathring{A}$, respectively
(the latter value corresponds to an adenine base). When DNA is present, $12$
additional counterions are added to the system.

For each instantaneous spatial configuration of the charges, the electrostatic
potential $\Phi(\mathbf{x})$ obeys the Poisson equation $\nabla\cdot
(\epsilon\nabla\Phi)=-4\pi\rho$, where the charge density $\rho\ $is obtained
by summation over ions $\rho(\mathbf{x})=\sum_{i}q_{i}S(\mathbf{x}%
-\mathbf{x}_{i})$, with $S$ being a smeared $\delta-$function (with the
boundary condition $\Phi=0$ on the surface of the box). This equation is
solved numerically using the conjugate-gradient method with grid spacings
$0.7$ $\mathring{A}$\ in the x,y-directions and $1.1$ $\mathring{A}$\ in the
z-direction. The force on the i-th particle is then given by $\mathbf{F}%
_{i}\mathbf{=-\nabla(}q_{i}\Phi\mathbf{+}U_{LJ}\mathbf{)+\Delta}\left(
q_{i}\sum_{j}q_{j}\mathbf{\hat{r}}_{ij}/\epsilon r_{ij}^{2}\right)  $ where
the last term accounts for the short-range correction to the electrostatic
force (the summation is taken over nearby ions within a Bjerrum length, $7$
$\mathring{A}$, and is computed in every time step). Since the rapid variation
of the electrostatic force between neighboring charges is taken into account
by this term, the remaining contribution to the electrostatic potential
($\Phi$) is calculated by solving the Poisson equation\ only once in every 10
time steps. The volume exclusion of particle cores is incorporated through the
Lennard-Jones potential $U_{LJ}(\mathbf{r}_{i})=\sum_{j}4\epsilon_{LJ}%
((\sigma/r_{ij})^{12}-(\sigma/r_{ij})^{6})$ for $r_{ij}\leq2^{1/6}\sigma$ and
$U_{LJ}=-\epsilon_{LJ}$ otherwise, where $\mathbf{r}_{ij}=\mathbf{r}%
_{i}-\mathbf{r}_{j}$ and $\sigma$ is the sum of the radii of the two
interacting particles. The temperature $T$ is used as the energy scale and
$\epsilon_{LJ}=k_{B}T$ is assumed for simplicity.

We first consider a nanopore without DNA. We run the simulation\ for $4000$
$ps$ (our time step corresponds to $3.4$ $fs$) and average the results over
time. Inspection of Table 1 shows that on the average there are about $2.2$
counterions and the same number of coions in the pore; however, the
fluctuations are of the same magnitude as the average. This value is lower
than that ($4$) corresponding to a uniform distribution of ions in the system,
in agreement with the expectation that ions are repelled from the pore because
the electrostatic self-energy of an ion in the pore is higher than in the
bulk, due to the low dielectric constant of the surrounding
membrane\cite{Parsegian}. In order to clarify the underlying physics, in
Fig.1a we plot time histograms of the numbers of counterions ($N_{+}$), coions
($N_{-}$) and the total charge ($Q$) in the pore. Notice that even though the
fluctuations around the mean values are quite large, these numbers fluctuate
nearly exactly in phase and therefore the total charge exhibits only small
variations ($\pm0.7e$) around zero. One is tempted to interpret the above
observation as evidence for formation of neutral Bjerrum (coion/counterion)
pairs\cite{Fowler} inside the pore and indeed, such pairs and even longer
string-like aggregates of alternating coions and counterions) are clearly
observable in snapshots of the instantaneous distribution of ions in the pore.

The time histograms of a pore with DNA\ inside are presented in Fig.1b. The
time averaged results for the numbers of counterions, coions and of DNA
charges in the pore are summarized in Table 1. In the physical case,
$\varepsilon_{m}=2$, inspection of Table 1 and Fig. 2a reveals that while the
counterion concentration in the pore reaches nearly $3$ times its value in the
bulk, coions are nearly completely depleted from it. The number of counterions
appears to be determined by the requirement that the negative charges of the
DNA are fully neutralized. Similar counterion enrichment and somewhat smaller
coion depletion are observed for abasic DNA (with bases removed) which has a
smaller excluded volume but the same charge as normal ssDNA.

In order to elucidate the role of membrane electrostatics on the distribution
of ions in the pore, we increased the dielectric constant of the membrane to
its value in water, $\varepsilon_{m}=80$. While a $4$-fold increase in the
concentration of coions in the pore compared to the $\varepsilon_{m}=2$ case
is observed, the number of counterions is nearly unchanged (see Table 1 and
Fig. 2b). We conclude that while the effect of increased electrostatic
self-energy of charges surrounded by low dielectric constant medium plays a
major role in coion depletion, the number of counterions in the pore is
determined mainly by the condition of electroneutrality. This leaves open the
question of whether the counterions are condensed (i.e., strongly localized)
on the DNA or free to move around the pore. While it is difficult to
distinguish between condensed and uncondensed counterions in a narrow pore,
inspection of Figs. 2a and 2b suggests that counterions are more strongly
localized on the DNA\ charges in the low $\varepsilon_{m}$ case. In order to
make a quantitative comparison we calculated the appropriate radial
distribution functions and found that the probability to find a counterion
within $4$ $\mathring{A}$ from a DNA charge is nearly $2$ times higher in the
$\varepsilon_{m}=2$ than in the $\varepsilon_{m}=80$ case. This is consistent
with the notion that counterion condensation on DNA is more pronounced in the
former case, because of the larger electrostatic self-energy of DNA in the
pore\cite{Parsegian}. Since stronger DNA-counterion coupling should lead to
stiffening of the macromolecule as the results of its \textquotedblleft
dressing\textquotedblright\ by the counterions, one expects the persistence
length of DNA\ to increase with decreasing $\varepsilon_{m}$. Indeed, we find
that the end-to-end distance is larger and that transverse fluctuations are
smaller when $\varepsilon_{m}=2$ than when $\varepsilon_{m}=80$ ($44$
$\mathring{A}$ versus $42$ $\mathring{A}$, respectively). Further evidence for
the renormalization of persistence length scenario comes from comparing the
end-to-end distances of abasic, uncharged (with charges removed from spheres
representing the phosphate groups) and normal ssDNA. As bases are added to the
charged\ backbone, $R$ increases from $35$ to $44$ $\mathring{A}$ and when
charges are removed from the latter $R$ decreases from $44$ to $40$
$\mathring{A}$, in accord with the expectation that the addition of the bulky
bases has a larger effect on the persistence length of DNA than its
\textquotedblleft dressing\textquotedblright\ by the small counterions. Since
nearly complete coion depletion from the pore is observed only when both (a)
DNA is present and (b) $\varepsilon_{m}=2$, it must be due to a combination of
excluded volume and electrostatic effects. A tentative explanation is that
while in the empty pore case Cl$^{-}$ ions can enter the pore by forming
Bjerrum pairs with K$^{+}$ ions, such bulky ion pairs are effectively excluded
from the pore in the presence of DNA. Since a similar mechanism should apply
even if the pore is occupied by a neutral macromolecule, we ran the simulation
with uncharged\ DNA. We find that there are about $1.4\pm1.0$ coions (and the
same number of counterions) in the pore, a number intermediate between that
corresponding to the empty channel ($2.2$) and the pore with a charged
macromolecule ($0.4$)\ cases. This again is consistent with the expectation
that the charged DNA is \textquotedblleft dressed\textquotedblright\ by the
neutralizing counterions which increase its excluded volume.

In summary, when ssDNA is present in the pore, it is neutralized by counterion
condensation; coions, on the other hand, are expelled from the channel. These
conclusions remain valid ($N_{+}=13,$ $N_{-}=1.5,$ $N_{-}^{DNA}=12,$ $R=39$
$\mathring{A}$) even if we increase the pore radius to $10$ $\mathring{A}$, a
value that is closer to the mean radius of an $\alpha$-hemolysin channel. In
spite of the various assumptions inherent in the present simulation (treating
the solution and the membrane as dielectric continua, neglecting hydration,
using a simplified model for DNA and taking no account of the detailed
molecular structure of the pore wall), the emerging physical picture of nearly
complete counterion condensation and coion depletion in the pore appears to be
quite reasonable. Analogy with electrophoresis suggests that the condensed
counterions (a) reduce the effective charge of DNA and (b) move together with
it in an electric field\cite{Bloomfield,Tanaka}. If this analogy is correct
one is tempted to conclude that these counterions do not take part in charge
transport across the pore and that the experimentally observed drop of the ion
current in the presence of DNA\cite{Meller2} should be attributed to the
reduction in the number of charge carriers due to DNA-induced depletion of
coions. However, since the identity of the charge carriers is sensitive to
molecular details of the pore neglected in our work (see, e.g., ref.
\cite{Roux}), the tentative conclusion that residual conduction\ through the
DNA-blocked pore is due to the negatively charged coions should be tested
experimentally e.g., by changing the chemical composition of the salt molecules.

Stimulating conversations with Amit Meller, Alexander Grosberg and Boris
Shklovskii are gratefully acknowledged. YR's work was supported by a grant
from the Israel Science Foundation and supported in part by the Institute for
Mathematics and its Applications with funds provided by the National Science
Foundation. MT's work and travel were supported by Grant-in-Aid No.16032217
from the Ministry of Education, Science and Culture of Japan. The present
computations were performed using supercomputers of the Minnesota University
Supercomputing Institute and the Institute of Molecular Science, Japan.

\newpage\noindent Table 1: The average numbers (and variances) of counterions
($N_{+}$), coions ($N_{-}$) and of DNA charges ($N_{-}^{DNA}$) in a pore of
$7.5\mathring{A}$ radius embedded in membranes of low ($\varepsilon_{m}=2$)
and high ($\varepsilon_{m}=80$) dielectric constants, are summarized. The net
charge in the pore in units of electron charge ($Q/e$) and the end-to-end
distance ($R$) of ssDNA within the pore are also shown. Only ions and DNA
charges with centers further than $1.8$ $\mathring{A}$ away from the ends of
the pore were counted in order to minimize end effects.%

\[%
\begin{tabular}
[c]{|c|c|c|c|c|}\hline
& empty ($\varepsilon_{m}=2$) & ss-DNA ($\varepsilon_{m}=2$) & ss-DNA
($\varepsilon_{m}=80$) & abasic-DNA ($\varepsilon_{m}=2$)\\\hline
$N_{+}$ & $2.2\pm1.5$ & $11.5\pm1.0$ & $11.8\pm1.2$ & $12.3\pm1.1$\\\hline
$N_{-}$ & $2.2\pm1.6$ & $0.4\pm0.5$ & $1.7\pm0.9$ & $0.9\pm0.8$\\\hline
$N_{-}^{DNA}$ & $--$ & $11.7\pm0.5$ & $11.8\pm0.4$ & $12.0\pm0.1$\\\hline
$Q/e$ & $0.06\pm0.7$ & $-0.6\pm0.9$ & $-1.7\pm1.2$ & $-0.6\pm0.7$\\\hline
$R$ & $--$ & $44\mathring{A}$ & $42\mathring{A}$ & $35\mathring{A}$\\\hline
\end{tabular}
\ \ \ \ \ \ \
\]



\newpage
\begin{figure}
\centerline{\scalebox{0.65}{\includegraphics{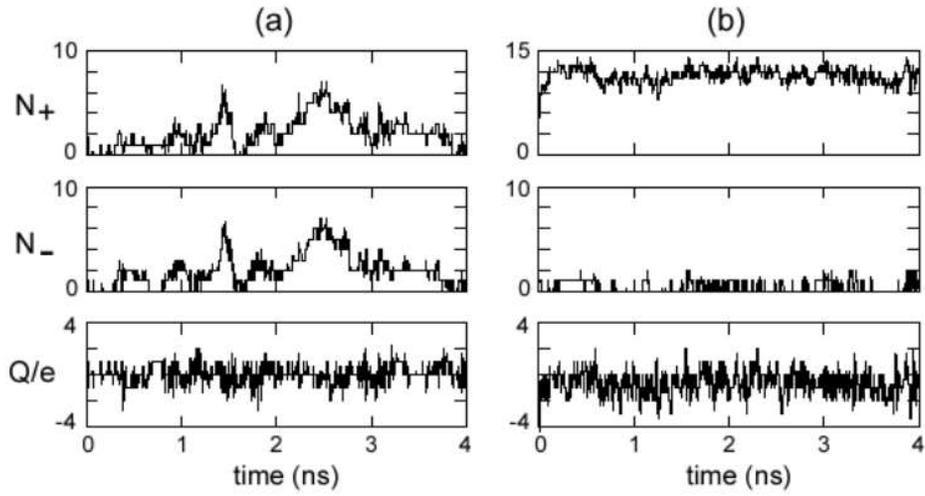}}}

\vspace*{0.5cm}
\caption{ 
 (a) The time history of the numbers of counterions
($N_{+}$), coions ($N_{-}$) and the total charges ($Q=e(N_{+}-N_{-})$)
contained in the empty pore (no DNA) is shown. (b) The time history in the
presence of DNA (with $N_{-}^{DNA}\approx12$) in the pore. The total
instantaneous charge in the pore is $Q=e(N_{+}-N_{-}-N_{-}^{DNA}).$ In both
cases $\varepsilon_{m}=2$.
}
\label{Fig.1}
\end{figure}

\newpage
\begin{figure}
\centerline{\scalebox{0.65}{\includegraphics{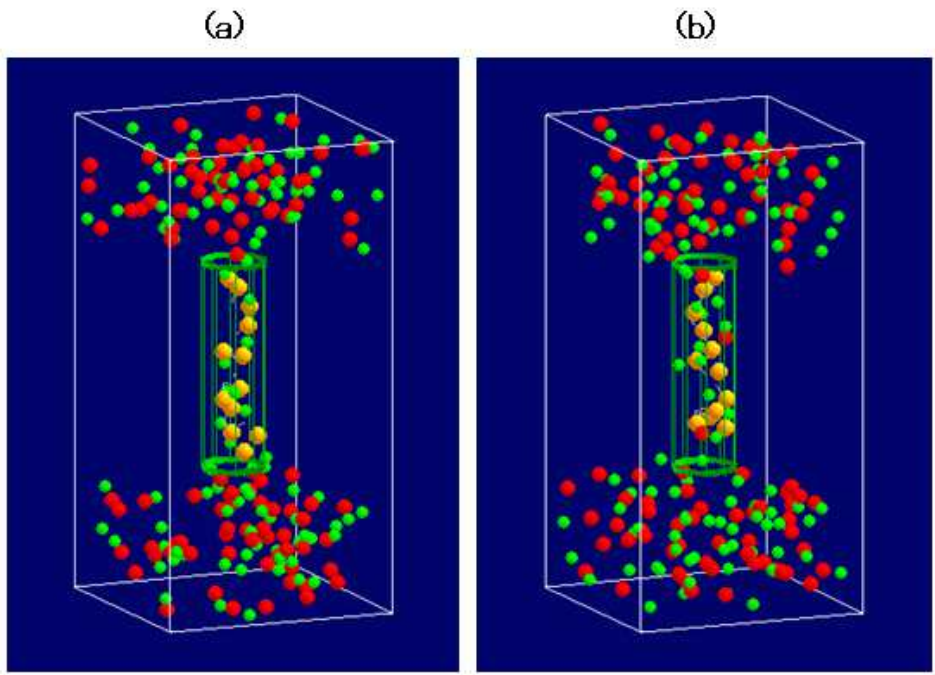}}}

\vspace*{0.5cm}
\caption{ 
Snapshot of the ion distribution with ssDNA in the pore
for $\varepsilon_{m}=2$ (a) and $\varepsilon_{m}=80$ (b). Green and red
spheres are counterions and coions, respectively and orange spheres represent
the charged phosphate groups of DNA (the neutral sugars and bases are not 
shown).}
\label{Fig.2}
\end{figure}

\end{document}